\documentstyle[12pt]{article}
\input amssym.def
\input amssym.tex

\newcommand{\qed}{\rule{3mm}{3mm}}

\newcommand{\cn}{{\rm cn}}
\newcommand{\dn}{{\rm dn}}
\newcommand{\sn}{{\rm sn}}

\newcommand{\rI}{{\rm I}}
\newcommand{\rJ}{{\rm J}}
\newcommand{\rM}{{\rm M}}
\newcommand{\rV}{{\rm V}}
\newcommand{\rW}{{\rm W}}

\newcommand{\bM}{{\bf M}}
\newcommand{\bV}{{\bf V}}
\newcommand{\bOm}{{\bf \Omega}}

\newcommand{\wM}{\widehat{\bM}}

\newcommand{\cV}{{\cal V}}

\newtheorem{theorem}{Theorem}
\newtheorem{proposition}{Proposition}
\newtheorem{lemma}{Lemma}
\newtheorem{definition}{Definition}

\begin{document}

\title{Integrable discretizations of the Euler top}
\author{A.I.\,Bobenko\footnote{E--mail: {\tt bobenko@sfb288.math.tu-berlin.de}}
\and B.\,Lorbeer\footnote{E--mail: {\tt lorbeer@math.tu-berlin.de}} 
\and Yu.B.\,Suris\footnote{E--mail: {\tt suris@sfb288.math.tu-berlin.de}}}
\date{Fachbereich Mathematik,
Technische Universit\"at Berlin, Str. 17 Juni 135, 10623 Berlin,
Germany}
\maketitle

\begin{abstract}
Discretizations of the Euler top sharing the integrals of motion with the 
continuous time system are studied. 
Those of them which are also Poisson with respect to the invariant 
Poisson bracket of the Euler top are characterized. For all these Poisson
discretizations a solution in terms of elliptic functions is found, allowing
a direct comparison with the continuous time case. We demonstrate that the
Veselov--Moser discretization also belongs to our family, and apply our
methods to this particular example.
\end{abstract}

\newpage

\section{Introduction}
The subject of integrable discretizations of integrable dynamical systems is 
slightly more than twenty years old. During the first decade of its existence
several different approaches were proposed for discretizing soliton equations 
\cite{AL}, \cite{H}, \cite{DJM}, \cite{QNCV}, \cite{NCW}, but neither of them 
dealt with integrable systems of the classical mechanics. The first examples
in this important subarea seem to appear about ten years ago \cite{V}, \cite{S}.
The Veselov's paper contained discrete time versions of such classical 
integrable systems as the Euler case of the rigid body motion and the Neumann
system. The algebraic construction behind these examples was later elaborated
in more detail in \cite{MV}. However, the nature of these exceptionally 
beautiful examples remains somewhat mysterious. They still resist to be included
in any of the existing general frameworks for integrable discretizations
(cf. \cite{DLT}). 
The origin of
the Veselov--Moser's discretizations, their place in the corresponding 
continuous time hierarchies, their relations to other existing examples,
and several further points remain to be clarified. The problem of comparing
the explicit solutions found in \cite{V}, \cite{MV} with the classical 
continuous time counterparts was also left open (despite the fact that on 
the level of equations of motion the continuous limit is easy to perform).

The present work may be considered as a sort of an extensive comment on 
\cite{V}, \cite{MV} in the part concerned with the Euler top. We do not 
close all the open problems mentioned above, but we do hope to bring some
light into some of them. In particular, we find formulas for the solution 
of the discrete time Euler top in the form which makes the comparison with
the continuous time case immediate. Moreover, the Veselov--Moser system
turns out to be by no means the only reasonable discretization of the
Euler top. We introduce a whole family of discretizations sharing integrals
of motion with the continuous time system, and characterize those of them
which share also the underlying invariant Poisson structure. In this
context the Veselov--Moser system becomes just one particular case, and not the
most simple one. 

It has to be mentioned that, unlike \cite{V}, \cite{MV}, our procedure for
obtaining explicit solutions is rather pedestrian; it does not use such
advanced tools as spectral theory of difference operators \cite{V} or 
Baker--Akhiezer functions \cite{MV}. Actually, our construction is based on the
addition formulas for elliptic functions, and could be invented
already by Jacobi. However, as a heuristic tool for finding it we used the
Lax representation of the Euler top in su(2) with a spectral parameter on 
an elliptic curve, which was certainly unknown in the times of Euler and Jacobi.

\section{The Euler top}

The famous Euler's equations describing the motion of a rigid body with the
fixed center of mass (Euler top) read \cite{G}:
\begin{eqnarray}\label{ET}
\dot{M}_1 & = & \left(\frac{1}{B}-\frac{1}{C}\right)M_2M_3\;,\nonumber\\
\dot{M}_2 & = & \left(\frac{1}{C}-\frac{1}{A}\right)M_3M_1\;,\\
\dot{M}_3 & = & \left(\frac{1}{A}-\frac{1}{B}\right)M_1M_2\;.\nonumber
\end{eqnarray}
Here $\bM=(M_1,M_2,M_3)^{\rm T}$ is the kinetic momentum vector in the 
coordinate system attached firmly to the body; the axes of this system coincide
with the principal axes of inertia, and the numbers $A,B,C>0$ are the 
corresponding moments of inertia. 

The vector form of the equations (\ref{ET}) is:
\begin{equation}\label{ET v}
\dot{\bM}=\bM\times\bOm(\bM)\;,
\end{equation}
where the vector of the angular velocity is introduced:
\begin{equation}\label{Omega}
\bOm(\bM)=\left(\frac{M_1}{A},\frac{M_2}{B},\frac{M_3}{C}\right)^{\rm T}\;.
\end{equation}

The equations (\ref{ET}) are Hamiltonian with respect to the following Poisson
bracket on ${\Bbb R}^3(\bM)$:
\begin{equation}\label{PB}
\{M_1,M_2\}=M_3\;,\qquad \{M_2,M_3\}=M_1\;,\qquad \{M_3,M_1\}=M_2\;.
\end{equation}
with the Hamilton function $H(\bM)=E(\bM)/2$, where
\begin{equation}\label{E}
E(\bM)=\frac{M_1^2}{A}+\frac{M_2^2}{B}+\frac{M_3^2}{C}\;.
\end{equation}
The bracket (\ref{PB}) is degenerate and has one Casimir function
\begin{equation}\label{M2}
M^2(\bM)=M_1^2+M_2^2+M_3^2\;.
\end{equation}
Generic symplectic leaves of the bracket (\ref{PB}) are two--dimensional 
spheres $M^2(\bM)={\rm const}$, hence each system Hamiltonian with respect to
this bracket is integrable in the Liouville--Arnold sense; in particular,
the Euler top is integrable.

An explicit solution to the equations (\ref{ET}) can be given in terms of the 
Jacobi elliptic functions \cite{G}. Suppose for the sake of definiteness that
\begin{equation}\label{ABC}
A>B>C>0\;,
\end{equation}
then
\[
\frac{1}{A}\leq\frac{E}{M^2}\leq\frac{1}{C}\;.
\]
The formulas for the solution look different depending on whether this quantity 
is greater or smaller than $1/B$:
\begin{equation}\label{sol}
M_1(t)=a\left\{\begin{array}{c} \cn\,\nu (t-t_0) \\ \dn\,\nu (t-t_0) 
\end{array}\right\},\quad M_2(t)=b\,\sn\,\nu (t-t_0),\quad
M_3(t)=c\left\{\begin{array}{c} \dn\,\nu (t-t_0) \\ \cn\,\nu (t-t_0) 
\end{array}\right\}.
\end{equation}
Here and below in similar situations the upper expressions in curly brackets 
refer to the case $\frac{1}{B}<\frac{E}{M^2}\leq\frac{1}{C}$, while the lower 
ones refer to the case $\frac{1}{A}\leq\frac{E}{M^2}<\frac{1}{B}$. The module 
$k$ of the elliptic functions in (\ref{sol}) is given by:
\begin{equation}\label{k}
k^2=\left\{\begin{array}{l}
\displaystyle\frac{(A-B)(M^2-CE)}{(B-C)(AE-M^2)} \\ \\
\displaystyle\frac{(B-C)(AE-M^2)}{(A-B)(M^2-CE)}
\end{array}\right\}\;,
\end{equation}
so that we have always $0<k^2<1$. The coefficients $a,b,c$ are defined by:
\begin{equation}\label{abc}
a^2=A\,\frac{M^2-CE}{A-C}\;,\quad
b^2=\left\{\begin{array}{c} B\,\displaystyle\frac{M^2-CE}{B-C} \\ \\
B\,\displaystyle\frac{AE-M^2}{A-B}
\end{array}\right\},\quad c^2=C\,\frac{AE-M^2}{A-C}\;.
\end{equation}
Finally, the frequency $\nu$ is given by:
\begin{equation}\label{nu}
\nu^2=\left\{\begin{array}{l}
\displaystyle\frac{(B-C)(AE-M^2)}{ABC} \\ \\
\displaystyle\frac{(A-B)(M^2-CE)}{ABC}
\end{array}\right\}\;.
\end{equation}
In the case $\frac{E}{M^2}=\frac{1}{B}$ the elliptic functions degenerate
to the hyperbolic ones, and we have:
\[
M_1(t)=\frac{a}{\cosh\,\nu (t-t_0)}\;,\quad 
M_2(t)=b\,\tanh\,\nu (t-t_0)\;,\quad
M_3(t)=\frac{c}{\cosh\,\nu (t-t_0)}\;,
\]
where
\[
a^2=\frac{A(B-C)}{B(A-C)}\,M^2\;,\quad b^2=M^2\;,\quad
c^2=\frac{C(A-B)}{B(A-C)}\,M^2\;,
\]
and
\[
\nu^2=\frac{(A-B)(B-C)}{AB^2C}\,M^2\;.
\]
In all three cases the numbers $a, b, c, \nu$ satisfy the condition
$abc\nu>0$. In what follows, we shall denote by $\nu(M^2,E)$ the
positive square root of the function given in (\ref{nu}), and hence
we must have $abc>0$.

\section{Discretizations}

Our aim here is to find {\it integrable discretizations} of the Euler top.
\begin{definition} An integrable discretization of the Euler top 
{\rm (\ref{ET v})}
is a one--parameter family of diffeomorphisms $F:\,{\Bbb R}^3\times
[0,\epsilon)\mapsto{\Bbb R}^3$,
\begin{equation}\label{F}
\wM=F(\bM,h)\;,
\end{equation}
such that each one of them $F(\cdot,h):\,{\Bbb R}^3\mapsto{\Bbb R}^3$
is Poisson with respect to the bracket {\rm(\ref{PB})}, has two integrals
of motion $M^2$ and $E$, i.e.
\begin{equation}\label{int}
M^2(F(\bM,h))=M^2(\bM)\;,\qquad E(F(\bM,h))=E(\bM)\;,
\end{equation}
and the following asymtotics hold:
\begin{equation}\label{as}
F(\bM,h)=\bM+h\bM\times\bOm(\bM)+o(h)\;, \quad h\to 0\;.
\end{equation}
\end{definition}

Actually, our maps always will be defined by {\it implicit} equations
of motion:
\begin{equation}\label{impl}
\wM-\bM=hf(\bM,\wM,h)\;.
\end{equation}

\begin{proposition} \label{impl map}
Let $f:\,{\Bbb R}^3\times{\Bbb R}^3\times[0,\epsilon)
\mapsto{\Bbb R}^3$ be a $C^1$--function in each argument such that
\[
f(\bM,\bM,0)=\bM\times\bOm(\bM)\;.
\]
Then for $h$ small enough the equation {\rm (\ref{impl})} defines a local 
diffeomorphism {\rm (\ref{F})} satisfying {\rm (\ref{as})}.
\end{proposition}
{\bf Proof.} This follows immediately from the implicit function theorem 
for all $h$ satisfying $\det(I-h\partial f/\partial \wM)\neq 0$, i.e. for all
 $h$ small enough. \qed
\vspace{1.5mm}

When speaking about discretizations, it is natural to think about $\bM$
in (\ref{F}) as about sequences $\bM_m:\,{\Bbb Z}\mapsto {\Bbb R}^3$ 
approximating solutions
$\bM(t):\,{\Bbb R}\mapsto {\Bbb R}^3$ of the Euler top (\ref{ET v}) in the 
sense that $\bM_m\approx \bM(mh)$. In this context the formula (\ref{F})
takes the form of the difference equation
\[
\bM_{m+1}=F(\bM_m,h)\;,
\]
and, similarly, the formula (\ref{impl}) has to be thought of as
\[
\bM_{m+1}-\bM_m=hf(\bM_m,\bM_{m+1},h)\;.
\]
The approximation property $\bM_m=\bM(mh)+O(h)$ holds on finite time intervals
and is assured by  (\ref{as}) (this is a standard fact from the numerical 
analysis).

\section{Elliptic coordinates and Poisson discretizations}

Now we would like to find a manageable criterium for a map (\ref{F}) to be
Poisson with respect to the bracket (\ref{PB}). The corresponding statement
becomes rather transparent in a new coordinate system in ${\Bbb R}^3(\bM)$.
The corresponding change of variables 
\[
\Phi:\,(M^2,E,\varphi)\mapsto (M_1,M_2,M_3)
\]
is suggested by the formulas (\ref{sol}) for a solution of the Euler top.
In the above formula $(M_1,M_2,M_3)\in{\Bbb R}^3$, and
$(M^2,E,\varphi)\in{\Bbb R}_+\times{\Bbb R}_+\times{\Bbb T}$, where
${\Bbb T}={\Bbb R}/(4K{\Bbb Z})$, and $K=K(k^2)$ is the full elliptic 
integral of the first kind corresponding to the value of $k^2$ given in
(\ref{k}). The formulas for $\Phi$ are different in two regions of
${\Bbb R}^3$ separated by the two planes
\begin{equation}\label{halfplanes}
\frac{E(\bM)}{M^2(\bM)}=\frac{1}{B}\quad\Leftrightarrow\quad
\left(\frac{1}{B}-\frac{1}{A}\right)M_1^2=
\left(\frac{1}{C}-\frac{1}{B}\right)M_3^2\;.
\end{equation}
Each one of the sets $\left\{\frac{E(\bM)}{M^2(\bM)}>\frac{1}{B}\right\}$
and $\left\{\frac{E(\bM)}{M^2(\bM)}<\frac{1}{B}\right\}$ consists of
two sectors, and each one of these sectors is bounded by two half-planes.

The {\it elliptic coordinates} in ${\Bbb R}^3(\bM)$ are introduced by:
\begin{equation}\label{ell coord}
M_1=a\left\{\begin{array}{c} \cn\,\varphi \\ \dn\,\varphi 
\end{array}\right\},\quad M_2=b\,\sn\,\varphi,\quad
M_3=c\left\{\begin{array}{c} \dn\,\varphi \\ \cn\,\varphi 
\end{array}\right\}.
\end{equation}
with $a,b,c$ given by (\ref{abc}), and the modulus $k^2$ of the elliptic 
functions given by (\ref{k}). On the subset $\left\{\frac{E}{M^2}>
\frac{1}{B}\right\}$ the sign of $c$ coincides with the sign of $M_3$
(which is constant in each one of two sectors), and the signs of $a,b$
satisfy the condition ${\rm sign}(ab)={\rm sign}(c)$. Similarly, on the
subset $\left\{\frac{E(\bM)}{M^2(\bM)}<\frac{1}{B}\right\}$ the sign of
$a$ coincides with the sign of $M_1$, while the signs of $b,c$ satisfy
${\rm sign}(bc)={\rm sign}(a)$. Finally, on the four half-planes described
by (\ref{halfplanes}) the elliptic coordinates are defined by continuity
according to
\[
M_1=\frac{a}{\cosh\,\varphi}\;,\quad 
M_2=b\,\tanh\,\varphi\;,\quad
M_3=\frac{c}{\cosh\,\varphi}\;,
\]
the signs of $a,c$ being the same as the signs of $M_1, M_3$, respectively,
and ${\rm sign}(b)={\rm sign}(ac)$.

\begin{proposition}
The formulas {\rm (\ref{ell coord})} define a valid change of variables 
(a local diffeomorphism) near each point of ${\Bbb R}^3$.
\end{proposition}
{\bf Proof.} It can be verified by a direct calculation that in the region
$\left\{\frac{E(\bM)}{M^2(\bM)}\neq\frac{1}{B}\right\}$ we have:
\[
\frac{\partial (M_1,M_2,M_3)}{\partial(M^2,E,\varphi)}=
-\frac{1}{4\nu(M^2,E)}\,\dn^2\,\varphi\;\neq\;0\;,
\]
and that the partial derivatives of which this Jacobian is composed allow
a continuation to the boundary $\left\{\frac{E(\bM)}{M^2(\bM)}=
\frac{1}{B}\right\}$. \qed

Now it is obvious that an arbitrary map (\ref{F}) having two integrals of
motion $M^2(\bM)$ and $E(\bM)$, can be cast, in the elliptic coordinates
$(M^2,E,\varphi)$, in the form
\begin{equation}\label{eq in ell}
\widehat{M}^2=M^2\;,\quad \widehat{E}=E\;,\quad \widehat{\varphi}=
\widehat{\varphi}(M^2,E,\varphi)\;.
\end{equation}
\begin{proposition}\label{symplectic}
The map {\rm (\ref{F})} having two integrals of motion $M^2(\bM)$ and $E(\bM)$
is Poisson with respect to the bracket {\rm (\ref{PB})} iff in the elliptic
coordinates $(M^2,E,\varphi)$ it takes the form
\begin{equation}\label{Poisson in ell}
\widehat{M}^2=M^2\;,\quad \widehat{E}=E\;,\quad \widehat{\varphi}=
\varphi+g(M^2,E)
\end{equation}
with the function $g$ not depending on $\varphi$.
\end{proposition}
{\bf Proof.} To prove this statement, we have to calculate the Poisson bracket
(\ref{PB}) in the coordinates $(M^2,E,\varphi)$. The corresponding formulas 
read:
\begin{equation}\label{PB in ell}
\{M^2,E\}=\{M^2,\varphi\}=0\;,\qquad \{E,\varphi\}=-2\nu(M^2,E)\;.
\end{equation}
Indeed, the function $M^2$ is a Casimir function, hence it Poisson commutes 
with both $E$ and $\varphi$. To calculate the bracket $\{E,\varphi\}$,
we substitute (away from the boundary $\frac{E}{M^2}=\frac{1}{B}$) the
expressions (\ref{ell coord}) into an arbitrary one of the formulas (\ref{PB}),
and after straightforward calculations arrive at the expression given above.

Actually, the concrete expression for $\{E,\varphi\}$ is not essential for
the proof of our proposition. The only important thing is that this Poisson 
bracket does not depend on $\varphi$. Indeed, we have:
\[
\{\widehat{E},\widehat{\varphi}\}=\frac{\partial\widehat{\varphi}}
{\partial\varphi}\,\{E,\varphi\}\;.
\]
Since the Poisson bracket $\{E,\varphi\}$ depends only on $M^2,E$ which are
integrals of motion, we see that the necessary and sufficient condition for
our map to be Poisson reads $\partial\widehat{\varphi}/\partial\varphi=1$,
which is equivalent to the last equation in (\ref{Poisson in ell}). \qed

\section{Special discretizations}
We derive now a family of discretizations of the Euler top. Our derivation is
based on a Lax representation with a spectral parameter for the Euler top.
We prefere to work with a ${\rm su}(2)$ Lax representation, since our experience
in discretizing various geometric structures (see, for example, \cite{BP})
conviced us that this procedure may be performed most straightforwardly when 
applied to ${\rm su}(2)$ Lax formulations.

To find a suitable Lax representation for the Euler top, we use a stationary
version of the Lax representation of the chiral field model due to Cherednik
\cite{Ch}.
Set 
\begin{equation}\label{2x2 M}
M(u)=\frac{1}{2i}\sum_{k=1}^3M_kw_k(u)\sigma_k\;,
\end{equation}
where $\sigma_k\; (k=1,2,3)$ are the Pauli matrices, and $w_k(u)$ are the 
following elliptic functions:
\begin{equation}\label{ws}
w_1(u) = \rho\,\frac{1}{\sn(u,\kappa)}\;,\quad
w_2(u) = \rho\,\frac{\dn(u,\kappa)}{\sn(u,\kappa)}\;,\quad
w_3(u) = \rho\,\frac{\cn(u,\kappa)}{\sn(u,\kappa)}\;.
\end{equation}
Here the parameter $\rho$ and the module $\kappa$ of the elliptic functions 
are defined by
\begin{equation}\label{rokappa}
\rho^2=\frac{A-C}{AC}\;,\qquad \kappa^2=\frac{C(A-B)}{B(A-C)}\;.
\end{equation}

Further, for a vector $\bV=(V_1,V_2,V_3)^{\rm T}\in{\Bbb R}^3$ set:
\begin{equation}\label{2x2 V}
V(u)=\frac{1}{2i}\sum_{k=1}^3V_kw_k(u-u_0)\sigma_k\;,
\end{equation}
where the point $u_0$ is chosen so that
\begin{equation}\label{u0}
w_1(u_0)=A^{-1/2},\qquad w_2(u_0)=B^{-1/2},\qquad w_3(u_0)=C^{-1/2}.
\end{equation}

Consider now the Lax equation
\begin{equation}\label{2x2 Lax}
\dot{M}(u)=[M(u),V(u)]\;.
\end{equation}
With the help of identities
\begin{equation}\label{ident 1}
w_j(u)w_k(u-u_0)=w_j(u_0)w_l(u-u_0)-w_k(u_0)w_l(u)\;,
\end{equation}
where $(j,k,l)$ is a permutation of $(1,2,3)$, one sees that the matrix equation
(\ref{2x2 Lax}) is equivalent to the set of the following two equations:
\begin{eqnarray}
\dot{\bM} & = & \bM\times\bOm^{1/2}(\bV)\;,  \label{gET 1}\\
0 & = & \bV\times\bOm^{1/2}(\bM)\;.          \label{gET 2}
\end{eqnarray}
Here the second equation is equivalent to $\bV=\gamma\,\bOm^{1/2}(\bM)$ with 
some $\gamma\in{\Bbb R}$, which, being substituted in the first equation, 
results in
\begin{equation}\label{gET}
\dot{\bM}=\gamma\,\bM\times\bOm(\bM)\;.
\end{equation}
Obviously, for a nonvanishing function $\gamma=\gamma(t)$ the latter 
equation is nothing but a time reparametrization of the Euler top
(\ref{ET v}). In this sense (\ref{2x2 Lax}) is a Lax representation 
of the Euler top.

Discretizing the above construction in time, we introduce the matrix
\begin{equation}\label{2x2 dV}
\cV(u)=I+\frac{h}{2i}\sum_{k=1}^3V_kw_k(u-u_0)\sigma_k\;,
\end{equation}
and consider instead of (\ref{2x2 Lax}) the discrete time Lax equation
\begin{equation}\label{2x2 dLax}
\widehat{M}(u)=\cV^{-1}(u)M(u)\cV(u)\;.
\end{equation}
(this equation may be interpreted as a stationary version of the lattice
chiral field model by \cite{NP}). 
Representing the latter equation as $\cV(u)\widehat{M}(u)=M(u)\cV(u)$ and using
the identities (\ref{ident 1}), we find that our matrix equation is equivalent
to the set of the following three equations:
\begin{eqnarray}
\wM-\bM & = & \frac{h}{2}\,(\bM+\wM)\times\bOm^{1/2}(\bV)\;,  \label{dgET 1}\\
0 & = & \bV\times\bOm^{1/2}(\bM+\wM)\;,          \label{dgET 2}
\end{eqnarray}
and 
\begin{equation}\label{dgET 3}
\sum_{k=1}^3(\widehat{M}_k-M_k)V_kw_k(u)w_k(u-u_0)=0\;.
\end{equation}
Now (\ref{dgET 2}) is equivalent to 
\begin{equation}\label{dgET aux}
\bV=\frac{1}{2}\gamma\,\bOm^{1/2}(\bM+\wM)
\end{equation}
with some $\gamma\in{\Bbb R}$, which, being substituted in (\ref{dgET 1}), 
results in
\begin{equation}\label{dgET}
\wM-\bM=\frac{1}{4}h\gamma\,(\bM+\wM)\times\bOm(\bM+\wM)\;.
\end{equation}
It remains to notice that, plugging (\ref{dgET aux}), (\ref{dgET}) into
(\ref{dgET 3}), we bring the latter equation to the form
\[
\sum_{k=1}^3 w_k(u)w_k(u-u_0)w_k(u_0)\Big(w_l^2(u_0)-w_j^2(u_0)\Big)=0\;,
\]
which is automatically satisfied due to the identity
\begin{equation}\label{ident 2}
w_l(u)w_l(u-u_0)w_j(u_0)-w_j(u)w_j(u-u_0)w_l(u_0)=w_k(u_0)\Big(w_l(u_0)
-w_j(u_0)\Big)\;.
\end{equation}
So, the matrix equation (\ref{2x2 dLax}) is equivalent to (\ref{dgET})
with some $\gamma\in{\Bbb R}$, the vector $\bV$ being given by (\ref{dgET
aux}).

However, the equation (\ref{dgET}) does not completely define the 
discretization due to arbitrariness of $\gamma$. In what follows we
shall consider this equation with $\gamma$ being a certain fixed function
on $\bM$, $\wM$. In other words, the subject of our further investigations
will consist of discretizations governed by implicit equations of
motion of the following special form:
\begin{equation}\label{impl spec}
\wM-\bM=\frac{1}{4}h\gamma(\bM,\wM,h)\,(\bM+\wM)\times\bOm(\bM+\wM)\;.
\end{equation}
We call them {\it special discretizations}. It is obvious that if 
\[
\gamma:\, {\Bbb R}^3\times{\Bbb R}^3\times [0,\epsilon)\mapsto{\Bbb R}_+
\]
is a $C^1$--function of each argument satisfying 
\begin{equation}\label{gamma0}
\gamma(\bM,\bM,0)=1\;,
\end{equation}
then the equation (\ref{impl spec}) fulfills the conditions of Proposition 
\ref{impl map}, and therefore defines a map (\ref{F}).

In components, special discretizations may be presented as:
\begin{eqnarray}
\widehat{M}_1-M_1 & = & \frac{1}{4}h\gamma(\bM,\wM,h)\,
\left(\frac{1}{B}-\frac{1}{C}\right)(M_2+\widehat{M}_2)(M_3+\widehat{M}_3)\;,
\label{impl spec 1}\\ \nonumber\\
\widehat{M}_2-M_2 & = & \frac{1}{4}h\gamma(\bM,\wM,h)\,
\left(\frac{1}{C}-\frac{1}{A}\right)(M_3+\widehat{M}_3)(M_1+\widehat{M}_1)\;,
\label{impl spec 2}\\ \nonumber\\
\widehat{M}_3-M_3 & = & \frac{1}{4}h\gamma(\bM,\wM,h)\,
\left(\frac{1}{A}-\frac{1}{B}\right)(M_1+\widehat{M}_1)(M_2+\widehat{M}_2)\;.
\label{impl spec 3}
\end{eqnarray}

\section{Poisson property of special discretizations}

We investigate now the integrability properties of special discretizations
(which are naturally expected due to the existence of a Lax representation
with a spectral parameter).
\begin{proposition}
Maps defined by equations of motion {\rm(\ref{impl spec})} possess
$M^2(\bM)$ and $E(\bM)$ as integrals of motion.
\end{proposition}
{\bf Proof.} It follows from (\ref{impl spec}) that
\begin{equation}\label{ints 1}
\langle\wM-\bM,\bM+\wM\rangle=0\quad {\rm and}\quad
\langle\wM-\bM,\bOm(\bM+\wM)\rangle=0\;,
\end{equation}
or, equivalently,
\begin{equation}\label{ints 2}
\langle\wM,\wM\rangle=\langle\bM,\bM\rangle\quad {\rm and}\quad
\langle\wM,\bOm(\wM)\rangle=\langle\bM,\bOm(\bM)\rangle\;.
\end{equation}
This proves our statement. \qed

Actually, this statement can be almost inverted. Namely, consider an arbitrary
discretization having $M^2(\bM)$ and $E(\bM)$ as integrals of motion. Then the
pairs $(\bM,\wM)$ satisfy (\ref{ints 2}), which is equivalent to (\ref{ints 1}).
If, in addition, $(\bM+\wM)\times\bOm(\bM+\wM)\neq 0$, then there exists a real
number $\gamma=\gamma(\bM,\wM)$ such that (\ref{impl spec}) holds. 

In what follows we shall need also the folowing technical and obvious
statement.
\begin{lemma}\label{technical}
The set of fixed points of the maps {\rm(\ref{F})} defined by {\rm(\ref{impl
spec})} coincides with the set of the points $\bM$ for which $(\bM+\wM)
\times\bOm(\bM+\wM)=0$, and this coincides with the union of the coordinate
axes
\[
\{M_1=M_2=0\}\cup\{M_2=M_3=0\}\cup\{M_3=M_1=0\}\;.
\]
For all other points $\bM$, at least two of the three expressions 
$M_1+\widehat{M}_1$, $M_2+\widehat{M}_2$, $M_3+\widehat{M}_3$ do not vanish.
\end{lemma}

Now we find conditions for a map defined by (\ref{impl spec}) to be Poisson.
To this end notice that the pairs $(\bM,\wM)=\Big(\bM,F(\bM,h)\Big)$ belong
to the subset of ${\Bbb R}^3(\bM)\times{\Bbb R}^3(\wM)$ singled out by the 
conditions $M^2(\bM)=M^2(\wM)$ and $E(\bM)=E(\wM)$. The elements of this subset 
may be parametrized by the quadruples $(M^2, E,\varphi,\widehat{\varphi})$ 
according to
\begin{equation}\label{M in ell}
\bM=\Big(a\left\{\begin{array}{c} \cn\,\varphi \\ \dn\,\varphi 
\end{array}\right\},\;b\,\sn\,\varphi,\; c\left\{\begin{array}{c} 
\dn\,\varphi \\ \cn\,\varphi \end{array}\right\}\Big)\;,
\end{equation}
\begin{equation}\label{wM in ell}
\wM=\Big(a\left\{\begin{array}{c} \cn\,\widehat{\varphi} \\ \dn\,
\widehat\varphi \end{array}\right\},\;b\,\sn\,\widehat\varphi,\; 
c\left\{\begin{array}{c} \dn\,\widehat\varphi \\ \cn\,\widehat\varphi 
\end{array}\right\}\Big)\;.
\end{equation}
Equivalently, we can use the following coordinates: $(M^2, E,\bar{\varphi},
\Delta\varphi)$, where
\begin{equation}\label{phis}
\bar{\varphi}=\frac{\varphi+\widehat{\varphi}}{2}\;,\qquad
\Delta\varphi=\frac{\widehat{\varphi}-\varphi}{2}\;.
\end{equation}
Let us denote on the above mentioned subset:
\begin{equation}\label{Gamma}
\gamma(\bM,\wM,h)=\Gamma(M^2,E,\bar{\varphi},\Delta\varphi,h)\;.
\end{equation}

\begin{theorem}\label{main} 
The equations of motion of the special discretization {\rm (\ref{impl spec})}
in the elliptic coordinates $(M^2,E,\varphi)$ have the form 
{\rm(\ref{eq in ell})}, where the function $\widehat{\varphi}(M^2,E,\varphi,h)$
is implicitly defined by the equation
\begin{equation}\label{master}
\frac{\Gamma(M^2,E,\bar{\varphi},\Delta\varphi,h)}
{1-k^2\sn^2(\bar{\varphi})\sn^2(\Delta\varphi)}=
\frac{2}{h\nu(M^2,E)}\,\frac{\sn(\Delta\varphi)}{\cn(\Delta\varphi)
\dn(\Delta\varphi)}
\end{equation}
The map defined by {\rm (\ref{impl spec})} is Poisson, iff the equation
{\rm(\ref{master})} may be solved for $\Delta\varphi$ as 
\begin{equation}\label{phi indep}
\Delta\varphi=\delta(M^2,E,h)=\frac{h\nu(M^2,E)}{2}\,\Big(1+O(h)\Big)
\end{equation}
with the function $\delta$ not depending on $\bar{\varphi}$.
\end{theorem}
{\bf Proof.} Denoting 
\begin{equation}\label{D}
D=D(M^2,E,\bar{\varphi},\Delta\varphi)=
1-k^2\sn^2(\bar{\varphi})\sn^2(\Delta\varphi)\;,
\end{equation}
we find from (\ref{M in ell}), (\ref{wM in ell}) with the help of addition 
formulae for elliptic functions:
\begin{equation}\label{M-M 1}
\widehat{M}_1-M_1=-\frac{2a}{D}\left\{\begin{array}{c}
\sn(\bar{\varphi})\dn(\bar{\varphi})\sn(\Delta\varphi)\dn(\Delta\varphi)\\
k^2\sn(\bar{\varphi})\cn(\bar{\varphi})\sn(\Delta\varphi)\cn(\Delta\varphi)
\end{array}\right\},\quad
\widehat{M}_1+M_1=\frac{2a}{D}\left\{\begin{array}{c}
\cn(\bar{\varphi})\cn(\Delta\varphi)\\ \dn(\bar{\varphi})\dn(\Delta\varphi)
\end{array}\right\},
\end{equation}
\begin{equation}\label{M-M 2}
\widehat{M}_2-M_2=\frac{2b}{D}\,\cn(\bar{\varphi})\dn(\bar{\varphi})
\sn(\Delta\varphi)\;,\quad
\widehat{M}_2+M_2=\frac{2b}{D}\,\sn(\bar{\varphi})
\cn(\Delta\varphi)\dn(\Delta\varphi)\;,
\end{equation}
\begin{equation}\label{M-M 3}
\widehat{M}_3-M_3=-\frac{2c}{D}\left\{\begin{array}{c}
k^2\sn(\bar{\varphi})\cn(\bar{\varphi})\sn(\Delta\varphi)\cn(\Delta\varphi)\\
\sn(\bar{\varphi})\dn(\bar{\varphi})\sn(\Delta\varphi)\dn(\Delta\varphi)\\
\end{array}\right\},\quad
\widehat{M}_3+M_3=\frac{2c}{D}\left\{\begin{array}{c}
\dn(\bar{\varphi})\dn(\Delta\varphi)\\ \cn(\bar{\varphi})\cn(\Delta\varphi)
\end{array}\right\}.
\end{equation}
Plugging this into an arbitrary one of the equations of motion 
(\ref{impl spec 1})-- (\ref{impl spec 3}), we arrive after
some cancellations at the equation (\ref{master}). Notice that Lemma 
\ref{technical} assures that, away from the fixed points, these cancellations 
are legitime in at least one of the equations of motion.

So, we have arrived at the equation (\ref{master}) of the form 
\begin{equation}\label{aux}
\Psi(M^2,E,\bar{\varphi},\Delta\varphi,h)=0\;,
\end{equation}
which serves for determining the function $\widehat{\varphi}(M^2,E,\varphi,h)$
for our map. The equation (\ref{aux}) may be rewritten as
\begin{equation}\label{aux1}
\Psi\Big(M^2,E,\frac{\varphi+\widehat{\varphi}}{2},
\frac{\widehat{\varphi}-\varphi}{2},h\Big)
\stackrel{{\rm def}}{=}\widetilde{\Psi}(M^2,E,\varphi,\widehat{\varphi},h)=0\;.
\end{equation}
Proposition \ref{symplectic} gives a necessary and sufficient condition 
$\partial\widehat{\varphi}/\partial\varphi=1$ for our map to be Poisson, and 
this is equivalent to the following condition:
\begin{equation}\label{aux2}
\frac{\partial\widetilde{\Psi}}{\partial\widehat{\varphi}}+
\frac{\partial\widetilde{\Psi}}{\partial\varphi}=0\quad {\rm on \;\;the\;\;
solutions\;\;of\;\;(\ref{aux1})}\;.
\end{equation}
Since, obviously,
\[
\frac{\partial\widetilde{\Psi}}{\partial\widehat{\varphi}}=
\frac{1}{2}\left(\frac{\partial\Psi}{\partial\bar{\varphi}}+
\frac{\partial\Psi}{\partial(\Delta\varphi)}\right)\;,\qquad
\frac{\partial\widetilde{\Psi}}{\partial\varphi}=
\frac{1}{2}\left(\frac{\partial\Psi}{\partial\bar{\varphi}}-
\frac{\partial\Psi}{\partial(\Delta\varphi)}\right)\;,
\]
we find that the Poisson property is equivalent to the following condition:
\begin{equation}\label{aux3}
\frac{\partial\Psi}{\partial\bar{\varphi}}=0\quad {\rm on \;\;the\;\;
solutions\;\;of\;\;(\ref{aux})}\;.
\end{equation}
In turn, (\ref{aux3}) assures that the solutions of (\ref{aux}) 
for $\Delta\varphi$ do not depend on $\bar{\varphi}$. \qed
\vspace{2mm}

{\bf Corollary.} In the conditions of the above theorem, solutions 
$\bM_m=(M_{1,m},M_{2,m},M_{3,m})^{\rm T}$ to the difference equation 
\begin{equation}\label{discr eq}
\bM_{m+1}-\bM_m=\frac{1}{4}h\gamma(\bM_m,\bM_{m+1},h)\,
(\bM_m+\bM_{m+1})\times\bOm(\bM_m+\bM_{m+1})
\end{equation}
in the integrable ($\equiv$ Poisson) case are given by
\begin{equation}\label{discr sol}
M_{1,m}=a\left\{\begin{array}{c} \cn(2m\delta+\varphi_0) \\ 
\dn(2m\delta+\varphi_0)
\end{array}\right\},\quad M_{2,m}=b\;\sn(2m\delta+\varphi_0),\quad
M_{3,m}=c\left\{\begin{array}{c} \dn(2m\delta+\varphi_0) \\ 
\cn(2m\delta+\varphi_0) 
\end{array}\right\}.
\end{equation}

\section{First examples of integrable discretizations}

We use Theorem \ref{main} to investigate the Poisson property of several
discretizations. We start with the following negative result.
\begin{proposition} The function $\gamma_0(\bM,\wM,h)\equiv 1$ defines a
non--Poisson map.
\end{proposition}
The simplest way to find Poisson (and therefore integrable) discretizations
is to assure that the left--hand side of the equation (\ref{master}) does
not depend on $\bar{\varphi}$.
\begin{proposition} The functions
\begin{equation}\label{gamma1}
\gamma_1(\bM,\wM,h)=\frac{4M^2}{\langle \bM+\wM,\bM+\wM\rangle}
\end{equation}
and
\begin{equation}\label{gamma2}
\gamma_2(\bM,\wM,h)=\frac{4E}{\langle \bM+\wM,\bOm(\bM+\wM)\rangle}
\end{equation}
define Poisson maps.
\end{proposition}
{\bf Proof.} Using the second expressions in (\ref{M-M 1})--(\ref{M-M 3}), 
it is not difficult to derive the following formulas:
\[
\frac{1}{4}\langle \bM+\wM,\bM+\wM\rangle=
\frac{a^2+c^2-b^2\sn^2(\Delta\varphi)}
{1-k^2\sn^2(\bar{\varphi})\sn^2(\Delta\varphi)}\;,
\]
\[
\frac{1}{4}\langle \bM+\wM,\bOm(\bM+\wM)\rangle=
\frac{\raisebox{2.5mm}{$\displaystyle\frac{a^2}{A}+\displaystyle\frac{c^2}{C}-
\displaystyle\frac{b^2}{B}\,\sn^2(\Delta\varphi)$}}
{1-k^2\sn^2(\bar{\varphi})\sn^2(\Delta\varphi)}\;.
\]
Hence, the left--hand side of the equation (\ref{master}) for 
$\gamma=\gamma_{1,2}$ does not depend on $\bar{\varphi}$. Taking into
account that, according to (\ref{abc}),
\[
a^2+c^2=M^2\;, \qquad \frac{a^2}{A}+\frac{c^2}{C}=E\;,
\]
we find the following expressions for this left--hand side:
\[
1-\alpha_{1,2}\sn^2(\Delta\varphi)\;,
\] 
where
\begin{equation}\label{alphas}
\alpha_1(M^2,E)=\frac{b^2}{M^2}\;,\qquad \alpha_2(M^2,E)=\frac{b^2}{BE}\;.
\end{equation}
(see (\ref{abc}) for the expression of $b^2$ through $M^2$ and $E$).
Therefore, the equation (\ref{master}) in these two cases is equivalent to
\begin{equation}\label{master12}
\frac{\sn(\Delta\varphi)}{\cn(\Delta\varphi)\dn(\Delta\varphi)}=
\frac{h\nu}{2}\,\frac{1}{1-\alpha_{1,2}\sn^2(\Delta\varphi)}\;.
\end{equation}
Obviously, its solutions satisfy
\[
\Delta\varphi=\delta_{1,2}(M^2,E,h)=\frac{h\nu(M^2,E)}{2}\,\Big(1+O(h)\Big)\;.
\]

\section{Veselov--Moser discretization}
In order to introduce the Veselov--Moser discretization, we need, first of all,
the matrix notation for the Euler top equation (\ref{ET v}). Using the
well--known isomorphism between the Lie algebra $\Big({\Bbb R}^3,\times\Big)$ 
with the Lie algebra $\Big({\rm so}(3),[\cdot,\cdot]\Big)$ of $3\times 3$
skew--symmetric matrices with the usual commutator, we can rewrite the equations 
of motion (\ref{ET v}) also in the matrix form
\begin{equation}\label{ET m}
\dot{\rM}=[\rM,\Omega(\rM)]\;,
\end{equation}
where
\begin{equation}\label{M m}
\rM=\left(\begin{array}{ccc} 0 & M_3 & -M_2 \\ -M_3 & 0 & M_1 \\
M_2 & -M_1 & 0\end{array}\right)
\end{equation}
and
\begin{equation}\label{Omega m}
\Omega(\rM)=\left(\begin{array}{ccc} 0 & M_3/C & -M_2/B \\ -M_3/C & 0 & M_1/A \\
M_2/B & -M_1/A & 0\end{array}\right)\;.
\end{equation}
The relation between the matrices $\rM$ and $\Omega=\Omega(\rM)$ may be 
expressed as follows:
\begin{equation}\label{M thru Omega}
\rM=\rJ\Omega+\Omega\rJ\;,
\end{equation}
where the entries of the diagonal matrix
\begin{equation}\label{J}
\rJ={\rm diag}(J_1,J_2,J_3)
\end{equation}
are defined by the relations
\begin{equation}\label{Js}
A=J_2+J_3\;,\qquad B=J_3+J_1\;,\qquad C=J_1+J_2\;.
\end{equation}
A spectral parameter dependent Lax representation for (\ref{ET m}) is:
\begin{equation}\label{Lax 3x3}
(\rM+\lambda\rJ^2)^{\raisebox{1.2mm}{\bf.}}=
[\rM+\lambda\rJ^2,\Omega(\rM)+\lambda\rJ]\;.
\end{equation}

Now we can describe the Veselov--Moser construction. The differential equation
(\ref{ET m}) is replaced by the difference one,
\begin{equation}\label{VM}
\widehat{\rM}=\omega^{\rm T}\rM\omega\;,
\end{equation}
where $\omega\in{\rm SO}(3)$ is an orthogonal matrix related to $\rM$ by 
means of the following relation, coming to replace, or to approximate, 
(\ref{M thru Omega}):
\begin{equation}\label{M thru omega}
h\rM=\omega\rJ-\rJ\omega^{\rm T}\;.
\end{equation}
It is easy to see that the previous two relations imply also
\begin{equation}\label{wM thru omega}
h\widehat{\rM}=\rJ\omega-\omega^{\rm T}\rJ\;.
\end{equation}
Moser and Veselov demonstrated that, in general, the equation 
(\ref{M thru omega}) does not determine the matrix $\omega\in{\rm SO}(3)$
uniquely. This non--uniqueness may be described as follows. Notice that
(\ref{M thru omega}), (\ref{wM thru omega}) are equivalent to the following
factorizations of matrix polynomials:
\[
\rI-h\lambda\rM-\lambda^2\rJ^2=(\omega+\lambda\rJ)(\omega^{\rm T}-\lambda\rJ)
\]
and
\[
\rI-h\lambda\widehat{\rM}-\lambda^2\rJ^2=(\omega^{\rm T}-\lambda\rJ)
(\omega+\lambda\rJ)\;,
\]
respectively.  Denote by $S$ the set of 6 roots of the equation
$\det(\rI-h\lambda\rM-\lambda^2\rJ^2)=0$. Then to each decomposition
$S=S_+\cap S_-$ into two disjoint sets of three roots satisfying
$S_+=\bar{S}_+=-S_-$ there corresponds a unique solution $\omega\in{\rm SO}(3)$
of (\ref{M thru omega}) such that $S_+$ is the set of the roots of
$\det(\omega+\lambda\rJ)=0$, while $S_-$ is the set of the roots of
$\det(\omega^{\rm T}-\lambda\rJ)=0$.

However, in the continuous limit, when $h$ is supposed to be small, there 
exists a {\it unique} decomposition, for which the roots from $S_-$
are positive real numbers $O(h)$--close to $J_1^{-1}$, $J_2^{-1}$, 
$J_3^{-1}$.
This is the only decomposition for which the corresponding $\omega$ has the 
following asymptotics:
\begin{equation}\label{omega as}
\omega=\rI+h\Omega(\rM)+O(h^2)\;.
\end{equation}
Supposing $h$ small enough, we consider from now on only this choice of 
$\omega$, and demonstrate that the discretization of Veselov and Moser also 
belongs to the class of special discretizations described by the equation 
(\ref{impl spec}).
\begin{theorem}
The Veselov--Moser discretization for $h$ small enough may be represented
in the form {\rm(\ref{impl spec})} with
\begin{equation}\label{gammaMV}
\gamma_{\rm VM}(\bM,\wM,h)=
\frac{2}{1+\sqrt{\,\raisebox{-1mm}{$1-\frac{1}{4}h^2||\bOm(\bM+\wM)||^2$}}}\;.
\end{equation}
\end{theorem}
{\bf Proof.} We shall need the following lemma, proof of which is put
in the Appendix.
\begin{lemma} For a matrix $\omega\in{\rm SO}(3)$, $O(h)$--close to $I$,
there exists a unique matrix $\rW\in{\rm so}(3)$ such that $\rW=O(h)$, and
\[
\omega=\rI+\rW+\frac{\gamma}{2}\rW^2\;,
\]
where
\[
\gamma=\frac{2}{1+\sqrt{1-|\rW|^2}}\;.
\]
Here
\[
|\rW|^2=W_1^2+W_2^2+W_3^2 \quad {\rm for} \quad 
\rW=\left(\begin{array}{ccc} 0 & W_3 & -W_2 \\ -W_3 & 0 & W_1\\
W_2 & -W_1 & 0\end{array}\right)\in{\rm so}(3)\;.
\]
\end{lemma}
With the help of this lemma we derive the following two equations:
\begin{equation}\label{MV aux1}
\omega-\omega^{\rm T}=2\rW\;,\qquad \omega+\omega^{\rm T}=2\rI+\gamma\rW^2\;.
\end{equation}
Now from (\ref{M thru omega}), (\ref{wM thru omega}) and the first equation
in (\ref{MV aux1}) we find:
\begin{equation}\label{MV aux2}
h(\rM+\widehat{\rM})=(\omega-\omega^{\rm T})\rJ+\rJ(\omega-\omega^{\rm T})
=2(\rW\rJ+\rJ\rW)\;,
\end{equation} 
hence
\begin{equation}\label{MV aux3}
\rW=\frac{1}{2}\,h\Omega(\rM+\widehat{\rM})\;.
\end{equation}
Further, from  (\ref{M thru omega}), (\ref{wM thru omega}) and the second 
equation in (\ref{MV aux1}) we find:
\begin{eqnarray}\label{MV aux4}
h(\widehat{\rM}-\rM) & = & \rJ(\omega+\omega^{\rm T})-(\omega+\omega^{\rm T})\rJ
=\gamma[\rJ,\rW^2]=\gamma[\rW\rJ+\rJ\rW,\rW]\nonumber\\
 & = & \frac{1}{4}\gamma h^2[\rM+\widehat{\rM},\Omega(\rM+\widehat{\rM})]\;,
\end{eqnarray} 
where
\[
\gamma=\frac{2}{1+\sqrt
{\,\raisebox{-1mm}{$1-\frac{1}{4}h^2|\Omega(\rM+\widehat{\rM})|^2$}}}\;.
\]
Using the above--mentioned isomorphism between $\Big({\Bbb R}^3,\times\Big)$ 
and $\Big({\rm so}(3),[\cdot,\cdot]\Big)$, we see that the theorem is proved.
\qed

It remains to reproduce from our point of view the result of Moser and Veselov 
concerning the Poisson property of their discretization. To this end, we 
demonstrate that the equation (\ref{master}) with $\Gamma_{\rm VM}$ 
corresponding to $\gamma_{\rm VM}$ allows a $\bar{\varphi}$--independent
solution for $\Delta\varphi$.
\begin{theorem} For the Veselov--Moser discretization the equation
{\rm(\ref{master})} is equivalent to
\begin{equation}\label{MV master}
\sn(\Delta\varphi)\cn(\Delta\varphi)\dn(\Delta\varphi)=\frac{h\nu}{2}\,\Big(1
+\alpha\,\sn^2(\Delta\varphi)-\beta\,\sn^4(\Delta\varphi)\Big)\;,
\end{equation}
where
\begin{equation}\label{MV alphabeta}
\alpha=\left\{\begin{array}{c}
\displaystyle\frac{2ACE-(A+C-B)M^2}{(B-C)(AE-M^2)} \\ \\
\displaystyle\frac{2ACE-(A+C-B)M^2}{(A-B)(M^2-CE)}\end{array}\right\}\;,\qquad
\beta= \left\{\begin{array}{c}
\displaystyle\frac{AC}{(B-C)^2}\,\displaystyle\frac{M^2-CE}{AE-M^2} \\ \\
\displaystyle\frac{AC}{(A-B)^2}\,\displaystyle\frac{AE-M^2}{M^2-CE}
\end{array}\right\}\;.
\end{equation}
The solution of {\rm(\ref{MV master})} satisfies
\begin{equation}\label{MV delta}
\Delta\varphi=\delta_{\rm VM}(M^2,E,h)=
\frac{h\nu(M^2,E)}{2}\,\Big(1+O(h)\Big)\;.
\end{equation}
\end{theorem}
{\bf Proof.} For the Veselov--Moser discretization the left--hand side of
the equation (\ref{master}) can be calculated with the help of (\ref{gammaMV})
and the second expressions in (\ref{M-M 1})--(\ref{M-M 3}):
\begin{equation}\label{GammaMV}
\frac{\Gamma(M^2,E,\bar{\varphi},\Delta\varphi,h)}
{1-k^2\sn^2(\bar{\varphi})\sn^2(\Delta\varphi)}=
\frac{2}{1-k^2\sn^2(\bar{\varphi})\sn^2(\Delta\varphi)+
\sqrt{\Big(1-k^2\sn^2(\bar{\varphi})\sn^2(\Delta\varphi)\Big)^2-h^2G}}\;,
\end{equation}
where
\begin{equation}\label{G}
G=\frac{a^2}{A^2}\left\{\begin{array}{c}
\cn^2(\bar{\varphi})\cn^2(\Delta\varphi) \\ 
\dn^2(\bar{\varphi})\dn^2(\Delta\varphi)\end{array}\right\}
+\frac{b^2}{B^2}\sn^2(\bar{\varphi})\cn^2(\Delta\varphi)\dn^2(\Delta\varphi)
+\frac{c^2}{C^2}\left\{\begin{array}{c}
\dn^2(\bar{\varphi})\dn^2(\Delta\varphi) \\ 
\cn^2(\bar{\varphi})\cn^2(\Delta\varphi)\end{array}\right\}\;.
\end{equation}
Hence the equation (\ref{master}) in the present case is equivalent to:
\[
1-k^2\sn^2(\bar{\varphi})\sn^2(\Delta\varphi)+
\sqrt{\Big(1-k^2\sn^2(\bar{\varphi})\sn^2(\Delta\varphi)\Big)^2-h^2G}=
h\nu\,\frac{\cn(\Delta\varphi)\dn(\Delta\varphi)}{\sn(\Delta\varphi)}
\]
We leave the radical alone on the left--hand side and square the resulting
equation to derive:
\begin{equation}\label{G aux}
-h^2G=h^2\nu^2\frac{\cn^2(\Delta\varphi)\dn^2(\Delta\varphi)}
{\sn^2(\Delta\varphi)}-2h\nu\,\frac{\cn(\Delta\varphi)\dn(\Delta\varphi)}
{\sn(\Delta\varphi)}\,\Big(1-k^2\sn^2(\bar{\varphi})\sn^2(\Delta\varphi)\Big)\;.
\end{equation}
Obviously, we have:
\begin{equation}\label{G01}
G=G_0(M^2,E,\Delta\varphi)-G_1(M^2,E,\Delta\varphi)\,\sn^2(\bar{\varphi})\;,
\end{equation}
where
\begin{equation}\label{G0}
G_0=\frac{a^2}{A^2}\left\{\begin{array}{c}
\cn^2(\Delta\varphi) \\ \dn^2(\Delta\varphi)\end{array}\right\}+
\frac{c^2}{C^2}\left\{\begin{array}{c}
\dn^2(\Delta\varphi) \\ \cn^2(\Delta\varphi)\end{array}\right\}\;,
\end{equation}
\begin{equation}\label{G1}
G_1=\frac{a^2}{A^2}\left\{\begin{array}{c}
\cn^2(\Delta\varphi) \\ k^2\dn^2(\Delta\varphi)\end{array}\right\}+
\frac{c^2}{C^2}\left\{\begin{array}{c}
k^2\dn^2(\Delta\varphi) \\ \cn^2(\Delta\varphi)\end{array}\right\}-
\frac{b^2}{B^2}\cn^2(\Delta\varphi)\dn^2(\Delta\varphi)\;.
\end{equation}
We shall prove that
\begin{equation}\label{G0G1}
G_1-k^2\sn^2(\Delta\varphi)G_0=
k^2\nu^2\cn^2(\Delta\varphi)\dn^2(\Delta\varphi)\;.
\end{equation}
This formula allows to derive from (\ref{G01}):
\begin{equation}\label{G aux2}
G=\frac{G_1}{k^2\sn^2(\Delta\varphi)}
\Big(1-k^2\sn^2(\bar{\varphi})\sn^2(\Delta\varphi)\Big)-
\nu^2\frac{\cn^2(\Delta\varphi)\dn^2(\Delta\varphi)}{\sn^2(\Delta\varphi)}\;,
\end{equation}
so that (\ref{G aux}) is equivalent to
\[
\frac{h^2G_1}{k^2\sn^2(\Delta\varphi)}=
2h\nu\,\frac{\cn(\Delta\varphi)\dn(\Delta\varphi)}{\sn(\Delta\varphi)}\;,
\]
or
\[
\sn(\Delta\varphi)\cn(\Delta\varphi)\dn(\Delta\varphi)=\frac{h}{2\nu k^2}G_1\;.
\]
This is equivalent to (\ref{MV master}), as one easily calculates from the
definition (\ref{G1}) that
\[
G_1=k^2\nu^2\Big(1+\alpha\,\sn^2(\Delta\varphi)-\beta\,\sn^4(\Delta\varphi)
\Big)
\]
with $\alpha$ and $\beta$ as in (\ref{MV alphabeta}). It remains to prove
the relation (\ref{G0G1}). But it follows from the definitions
(\ref{G0}), (\ref{G1}):
\[
G_1-k^2\sn^2(\Delta\varphi)G_0=
\left\{\begin{array}{c}\displaystyle\frac{a^2}{A^2}+
\displaystyle\frac{k^2c^2}{C^2}-\displaystyle\frac{b^2}{B^2} \\ \\
\displaystyle\frac{k^2a^2}{A^2}+
\displaystyle\frac{c^2}{C^2}-\displaystyle\frac{b^2}{B^2}\end{array}\right\}
\cn^2(\Delta\varphi)\dn^2(\Delta\varphi)\;,
\]
and the expressions in the curly brackets here in both cases are equal to
$k^2\nu^2$, due to (\ref{abc}), (\ref{k}), (\ref{nu}). \qed

\section{Conclusions}
In the present work we have introduced a large family of discretizations
of the Euler top sharing the integrals of motion with the continuous time 
system. We characterized those of them which are also Poisson with respect to
the invariant Poisson bracket of the Euler top. For all these Poisson
discretizations we found a solution in terms of elliptic functions which allows
a direct comparison with the continuous time case. We demonstrated that the
Veselov--Moser discretization also belong to our family, and applied our
methods to this particular example.

Let us mention some of the possible dirsections of the further progress.
First of all, our construction works on the level of reduced equations
of motion of the rigid body (in the Lie algebra ${\Bbb R}^3\approx{\rm so}(3)
\approx{\rm su}(2)$).
It would be important to lift it to the level of the cooresponding Lie group
SO(3) (or SU(2)). On a more mechanical language, we want to discretize the 
equations of motion in the rest frame, and not only in the frame attached 
firmly to the body. Notice that the Veselov--Moser construction has a 
variational
(Lagrangian) origin, which makes it valid in the group. Another example of such 
a discretization admitting a variational formulation in the corresponding group 
is the recently found discrete time Lagrange top \cite{B}.

As a second problem, our construction has to be generalised to higher dimensions
(for the general Euler--Manakov case of the multidimensional rigid body).
Notice that the Veselov--Moser construction goes through in higher dimensions.

Futher, it would be important to include our construction into the general
$r$--matrix framework, connected to factorizations in the loop groups,
cf. \cite{S2}.
 

\begin{appendix}
\section{Proof of Lemma 2} 
Let $\omega\in{\rm SO}(3)$, $\omega=\rI+O(h)$. Set $\rV=\displaystyle\frac
{\omega-\rI}{\omega+\rI}$, then $\rV\in{\rm so}(3)$, $\rV=O(h)$, and
\[
\omega=\frac{\rI+\rV}{\rI-\rV}=\frac{2}{\rI-\rV}-\rI\;.
\]
But for $\rV\in{\rm so}(3)$ we have: $\rV^3=-|\rV|^2\rV$, and by induction
\[
\rV^{2k+1}=(-1)^k|\rV|^{2k}\rV\;,\quad \rV^{2k+2}=(-1)^k|\rV|^{2k}\rV^2\;,
\quad k\ge 1\;,
\]
so that
\[
\frac{\rI}{\rI-\rV}=\rI+\frac{1}{1+|\rV|^2}\,\rV+\frac{1}{1+|\rV|^2}\,\rV^2\;,
\]
hence
\[
\omega=\rI+\frac{2}{1+|\rV|^2}\,\rV+\frac{2}{1+|\rV|^2}\,\rV^2\;.
\]
Now setting
\[
\rW=\frac{2}{1+|\rV|^2}\,\rV\;,
\]
so that $\rW\in{\rm so}(3)$, $\rW=O(h)$, we find:
\[
\omega=\rI+\rW+\frac{\gamma}{2}\,\rW^2\;,\quad {\rm where}\quad 
\gamma=1+|\rV|^2\;.
\]
We have, obviously, 
\[
|\rW|^2=\frac{4}{\left(1+|\rV|^2\right)^2}\,|\rV|^2\;\;\Longrightarrow\;\;
|\rV|^2=\frac{2-|\rW|^2-2\sqrt{1-|\rW|^2}}{|\rW|^2}\;,
\]
and
\[
\gamma=1+|\rV|^2=\frac{2-2\sqrt{1-|\rW|^2}}{|\rW|^2}=\frac{2}{1+
\sqrt{1-|\rW|^2}}\;.
\]
This proves the lemma. \qed
\end{appendix}

\begin{thebibliography}{WWW}

\bibitem[AL]{AL}  M.Ablowitz, J.Ladik.  A nonlinear difference scheme and
  inverse scattering. {\em Stud. Appl. Math.}  {\bf 55} (1976) 213--229;
 On solution of a class of nonlinear partial difference equations. 
{\em Stud. Appl. Math.} {\bf 57} (1977) 1--12.

\bibitem[H]{H} R.Hirota. Nonlinear partial difference equations. I--V. 
{\em J. Phys. Soc. Japan} {\bf 43} (1977)  1423--1433, 2074--2078, 2079--2086;
{\bf 45} (1978) 321--32; {\bf 46} (1978) 312--9.

\bibitem[DJM]{DJM} F.Date, M.Jimbo,  and T.Miwa. Method for generating 
discrete soliton equations. I--IV. {\em J. Phys. Soc. Japan} {\bf 51} (1982)
4116--4124, 4125--4131; {\bf 52} (1983) 761--765, 766--771.

\bibitem[QNCV]{QNCV}  G.Quispel, F.Nijhoff, H.Capel,  and J.Van der Linden 
Linear integral equations and nonlinear differential--difference equations. 
{\em Physica A}   {\bf 125} (1984) 344--380.

\bibitem[NCW]{NCW} F.W.Nijhoff, H.Capel, and G.Wiersma. Integrable lattice 
systems in two  and three dimensions. {\em Lect. Notes Phys.} {\bf 239} (1984)  
263--302; Linearizing integral transform for the multicomponent lattice KP. 
{\em Physica A} {\bf 138} (1986) 76--99.

\bibitem[V]{V} A.P.Veselov.  Integrable systems with discrete time and 
difference operators. {\em Funct. Anal. Appl.} {\bf 22} (1988) 1--13.

\bibitem[S]{S} Yu.B.Suris.   Generalized Toda chains in discrete time. 
{\em Algebra i Anal.} {\bf 2} (1990) 141--157; Discrete--time generalized 
Toda lattices: complete integrability and relation with relativistic Toda 
lattices. {\em Phys. Lett. A} {\bf 145} (1990) 113--119.

\bibitem[MV]{MV} J.Moser, A.P.Veselov.  Discrete versions of some 
classical integrable systems and factorization of matrix polynomials. 
{\em Commun. Math. Phys.} {\bf 139} (1991) 217--243.

\bibitem[DLT]{DLT} P.Deift, L.-Ch.Li,  and C.Tomei. Loop groups, discrete
versions of some classical integrable systems, and rank 2 extensions.
{\em Mem. Amer. Math. Soc.} {\bf 479} (1992).

\bibitem[G]{G} V.V.Golubev. Lectures on the integration of the equations of
motion of a rigid body about a fixed point. State Publishing, Moscow, 1953. 

\bibitem[BP]{BP} A.I.Bobenko, U.Pinkall. Discretization of surfaces and
integrable systems. -- In: Discrete Integrable geometry and Physics, Eds.  
A.Bobenko and R.Seiler, Oxford University Press, 1988. 

\bibitem[Ch]{Ch} I.V.Cherednik. On the integrability of the equation of a 
two--dimensional asymmetric chiral O(3)--field and its quantum anlogue.
{\em Yadernaya Fiz.} {\bf 33} (1981) 278--282. Engl. transl. in {\em Sov. J.
Nucl. Phys.} {\bf 33} (1981).

\bibitem[NP]{NP} F.W.Nijhoff, V.Papageorgiou. Lattice equations assosiated with
the Landau--Lifshitz equations. {\em Phys. Lett. A} {\bf 141} (1989) 269--274.

\bibitem[B]{B} A.I.Bobenko. Discrete integrable systems and geometry.
{\em Proc. Int. Congress Math. Phys. '97} (to appear).

\bibitem[S2]{S2} Yu.B.Suris. $R$--matrices and integrable discretizations. --
In: Discrete Integrable geometry and Physics, Eds.  A.Bobenko and R.Seiler,
Oxford University Press, 1988. 
\end{thebibliography}
\end{document}